\documentstyle[12pt]{article}
\begin{document}

\def\lash#1{#1\!\!\!  / } \def\half{{1\over 2}} \def\ml{m_L} 
\def\ms{m_S} \def\gl{\Gamma_L} \def\dh{\lash{\delta h}} \def\ko{K_0} 
\def\kob{{\overline K}_0} \def\kl{K_L} \def\ks{K_S} \def\rk{\rho_K} 
\def\rkk{\rho_{KK}} \def\rl{\rho_L} \def\rs{\rho_S} \def\ri{\rho_I} 
\def\rib{\rho_{\bar I}} \def\eps{\epsilon_{S}} \def\epl{\epsilon_{L}} 
\def\alphab{{\overline \alpha}} \def\betab{{\overline \beta}} 
\def\gammab{{\overline \gamma}}
\def\etapp{\eta_{+-}} \def\gs{\Gamma_S} \def\pp{\pi\pi} \def\ppp{3\pi} 
\def\pln{\pi\ell\nu} \def\rpp{R_{\pi\pi}}
\def\gam{\Gamma} \def\dm{\Delta m} \def\dg{\Delta \gam} \def\gl{\gam_L} 
\def\gs{\gam_S} \def\\ml{m_L} \def\ms{m_S} \def\gb{{\bar \gam}} 
\def\mb{{\bar m}} \def\egl{e^{-\gl\tau}} \def\egs{e^{-\gs\tau}} 
\def\egi{e^{-\gb\tau}} \def\egic{e^{-(\gb+\alpha-\gamma)\tau}} 
\def\em{e^{-i\dm\tau}} \def\emb{e^{+i\dm\tau}} 
\def\plm{\pi^{\pm}\ell^{\mp}\nu} \def\lp{\pi^{-}\ell^{+}\nu} 
\def\lm{\pi^{+}\ell^{-}\nu}
\def\phipp{\phi_{+-}}
\def\P{{\cal P}} \def\PP{\bar{\cal P}} \def\Q{{\cal Q}}
\def\Dt{\Delta\tau} \def\O{{\cal O}} \def\M{{\cal M}}
\def\ra{\rightarrow} \def\Re{{\rm Re}} \def\Im{{\rm Im}}
\def\qmg{{(\diamondsuit)}} \def\rrho{{\hat{\rho}}}
\def\ket#1{|#1\rangle} \def\bra#1{\langle#1|}
\def\VEV#1{\langle#1\rangle}
\def\cpt{{\it CPT\/}}
\def\cptc{\cpt\ conservation} \def\cptv{\cpt\ violation} \def\cp{{\it
CP\/}} \def\cpv{\cp\ violation} \def\cpc{\cp\ conservation}
\def\phif{$\phi\,$ factory} \def\phifs{$\phi\,$ factories}
\def\qm{quantum mechanics} \def\qml{quantum mechanical}
\def\qft{quantum field theory}
\def\kkbar{$\ko$--$\kob$}
\rightline{UW/PT 96 -- 11}
\rightline{ July 25, 1996}

\vskip 4.5cm

\begin{center}
 {CPT VIOLATION FROM PLANCK SCALE PHYSICS\footnote{Talk presented at
the Workshop on K physics,  Orsay, France, May 30 - June 4, 1996.}}
\end{center} 
\medskip
\begin{center}
	Patrick Huet\\
	Department of Physics, Box 351560 \\
	University of Washington\\
	Seattle  WA 98195
\end{center}
\bigskip
\begin{center}
 \parbox[]{\textwidth}{
 \noindent I address the phenomenology of \cptv\ in the neutral kaon system 
 under the assumption that it originates from Planck scale physics.  
 This assumption opens the door to a new set of \cpt\ violating 
 parameters whose phenomenology is distinct from the $\Delta$ 
 parameter usually considered in the Hamiltonian.  The origin of these 
 parameters reflects a possible departure from a $S$-matrix evolution.  
 Existing bounds on \cptv\ are near the expected 
 range based on naive dimensional analysis.  This provides a strong 
 incentive to pursue the quest of \cptv\ in near-future kaon 
 experiments.
}
\end{center}

\vfill
\newpage
 
In 1983, S.W. Hawking${}^{1]}$ extrapolated on earlier developments
in the quantum theory of gravity, and proposed a generalization of \qm\
which allows the evolution of pure states to mixed states. This
generalization of \qm\ departs from a $S$-matrix evolution and was
shown to conflict\footnote{ \cpt\
conservation could be preserved in a weaker form.} with \cpt\
conservation.${}^{2]}$ It is
worth noting that departure from a unitary evolution predicted by \qm\ 
had already been given an experimental scrutiny in the early 
70's,${}^{3]}$ without reference to a particular theoretical framework.

Soon after, Ellis, Hagelin, Nanopoulos, and Srednicki${}^{4]}$
observed that systems which possesses a high degree of quantum
coherence are most appropriate to probe the violation of \qm\ of the
type proposed by Hawking.  They conveniently wrote a differential form
of Hawking's generalization of the time evolution of a density matrix
$\rk$, namely,

\begin{equation}   
i {d\over d\tau} \rk =    H\,\rk - \rk\, H^{\dagger}
\,+\,\lash{\delta h} \,\rk .
\label{eq:densitytime}
\end{equation}

The first two terms on the RHS of this equation, accounts for the quantum 
mechanical evolution of the system, which is described by an 
Hamiltonian $H$. The third term accounts for the loss of coherence in 
the evolution of the beam; $\dh$ is a linear operator and is written 
so to require that it does not break conservation of probability 
and does not decrease the entropy of the system.

One of the simplest systems exhibiting ``macroscopic" quantum
coherence is a beam of neutral kaons.  As it so happens, a major
experimental consequence of the presence of the $\dh$--term in the
\kkbar\ system is \cptv. There is no a priori reason for \cpt\
symmetry to be exactly conserved.  \cptc, however, arises naturally in
the framework of a local, Lorentz-invariant, \qft --- the tool of
modern theoretical physics.  The immense success of the latter is the
root of the conjecture that \cpt\ may be exactly conserved.  Previous
experimental searches for \cptv\ in departing for that conjecture,
have assumed $\dh = 0$.  This is the reason why \cptv\ has been
parameterized with a single (complex) parameter $\Delta, \, \, = (\eps
-\epl)/2$, in the effective Hamiltonian $H$, reflecting a difference
in mass or in decay properties between $\ko$ and $\kob$.  One should
remember, however, that the motivation for the \cpt\ violating
perturbation $\dh$ arises from considering quantum gravitational
effects, i.e., quantum fluctuations of the spacetime, which, until
today, have not been successfully incorporated in a \qft.\footnote{ A
low energy effective field theory which successfully describes aspects
of quantum gravity, cannot insure \cptc\ as it does not describe short
distances physics.}  In the future, the latter might turn out to
describe those fluctuations, or, perhaps more likely, a more general
(yet to be developed) framework might be required to meet the
challenge of dealing with a ``spacetime foam".  That framework will
not be required to conserve \cpt\ symmetry.  In fact, preliminary
theoretical considerations suggest that it might not.  Some of these
considerations are: $(a)$ Quantization of matter in presence of a
gravitational background; the extrapolation of which suggests the
introduction of the \cpt--violating term $\dh$.  $(b)$ Local, but
non-Lorentz-invariant, operators induced in the low energy world by
the intrinsic non-local Planckian structure of string theory;${}^{5]}$
these operators induce a non-zero value for the \cpt\ violating
parameter $\Delta$ (cf.  Alan Kosteleck\'y's contribution to these
proceedings).\footnote{For more speculative consequences of string
theory leading to the generation of a $\dh$ term, see  John Ellis'
contribution to these proceedings and references therein.}

At the light of these speculative, yet suggestive, theoretical
considerations, it is appropriate to seriously consider Planck scale
physics as a prime candidate for inducing \cptv\ and to generalize the
search of \cptv\ in the \kkbar\ system from the single parameter
$\Delta$, to the more complete set $\{ \Delta, \dh^{{ij}}/\dm \}$ (I
have appropriately divided $\dh^{ij}$ with the mass difference $\dm$,
to obtain a dimensionless parameter).  This addition has more than an
academic value since these parameters have a different experimental
signature --- a nice feature when it comes to extracting them from
data.${}^{6]}$

The assumption of Planckian physics as the origin of \cptv\ has also
the virtue of providing a scale which allows to make an
order-of-magnitude estimate and so to give experimentalists a range to
shoot for.  One naively expects

\begin{equation}
\Delta \ \sim \ {\dh \over \Delta M} \ \sim \ {E \over \Delta M} \ \left( 
{E \over M_{\rm Pl} } \right)^{n} \, .
\label{estimate}
\end{equation}
Here, $E$ is the energy of the system (say, the kaon mass), $1 / 
\dm$ is the typical coherence time of the system (say the 
oscillation length of the kaon), and $n$ is an unknown 
power.\footnote{ Although, it has been suggested in this workshop that 
$n$ could assume arbitrary non-integer values, I will restrict it to 
near-integer values as otherwise this exercise in dimensional analysis 
looses its interest.} In the particular case of the kaon system, we 
estimate $\{ \Delta, \dh^{{ij}}/\dm \} \sim 10^{-5}$ for $n=1$; that 
is, about $1 \% $ of the total \cp\ violation observed in the kaon 
system.  This is precisely near the range accessible by current experiments.  
Higher values of $n$ renders the effect nearly inobservable.  These 
estimates, taken at faith value, are very suggestive: \cptv\ is at the 
reach of near-future experiments, otherwise, it is unlikely to be seen 
at all unless it originates from physics at a scale far below $M_{\rm 
Pl}$.

Let us take a closer look at a beam of kaons evolving under Eq.
(\ref{eq:densitytime}) and understand how \cptv\ originates in this
particular system.  The remaining of this paper follows closely the
analysis of Ref.~\cite{hp} (See Ref.~\cite{ellisetal} for an
alternative analysis).


\section*{Violation of \cpt\ in the \kkbar\ system}
Any observable $\VEV{{\cal P}}$ along the kaon beam can be computed by
tracing the product of the density matrix $\rk(\tau)$ with an
appropriate operator $\O_ \P$, as $\VEV{{\cal P}} = {\rm Tr} \bigl[
\rk \O_\P \bigr]$.  The time evolution of $\rk(\tau)$ is determined by
Eq.~(\ref{eq:densitytime}) and is completely characterized by an
effective Hamiltonian $ H,\,\, = M - {i\over 2} \Gamma$, which
incorporates the natural width of the system,\footnote{as well as \cp\
and \cpt\ perturbations compatible with \qm.} and by a linear operator
$\dh$.  The latter is only constrained so as not to break conservation of
probability and not to decrease the entropy of the system; that makes
it expressible in terms of six parameters.  In order to lower this
number to a more tractable one, one neglects its strangeness violating
components, reducing $\dh$ to three unknown positive parameters
written as $\alphab \cdot \dm$, $\betab \cdot \dm$ and $\gammab \cdot
\dm$.  They satisfy the relation $\alphab \gammab > \betab^2$, they
are dimensionless and might be as large as $m_K^2/(\dm \, m_{\rm Pl})
\sim 3 \ 10^{-5}$.  These are the parameters which eventually shift
the observed value of the \cp\ violating parameter $\epsilon \sim 2.32 \times
10^{-3}$.  The solution of
Eq.~(\ref{eq:densitytime}) is generally expressible as

\begin{equation}\rk(\tau)\,=\,A_L \rl^\qmg 
\egl+A_S\rs^\qmg \egs+\left(\, A_I\ri^\qmg \egi\em+  \ {\rm h.c.} \ \right)
\label{eq:rhotau}
\end{equation} 

The parameters $A_{S,L,I}$ are fixed by the production mechanism of
the beam.  In the absence of the \qm\ violating perturbation
$\dh\,\rk$ in Eq.~(\ref{eq:densitytime}), the eigenmodes $\rl$, $\rs$
and $\ri$, are expressible in
terms of the pure states $\ket{\kl}$ and $\ket{\ks}$ as
$\rl^\qmg\,=\,\ket{\kl}\bra{\kl}$, $\rs^\qmg\,=\,\ket{\ks}\bra{\ks}$
and $\ri^\qmg\,=\, \ket{\ks}\bra{\kl}$, while $\gb= (\gl + \gs)/ 2$,
$\dg= \gs - \gl$ and $\dm = \ml-\ms$.  After adding the \qm\ violating
term $\dh \,\rk$, the eigenmodes are changed to, in first order in
small quantities ($\dm+i\dg/2 \simeq i \sqrt{2} \dm \, e^{-i\phi_{\rm SW}}$)
 
\begin{eqnarray}
\rl= \rl^\qmg +  {\gammab \over 2}  \rs^\qmg &-&  \, {\betab \over
\sqrt{2}} \left(\, i e^{i\phi_{SW}} \ri^\qmg + \ {\rm h.c.} \ \right) \
, \
\rs= \rs^\qmg -  {\gammab \over 2}  \rl^\qmg \ - \  \, {\betab \over
\sqrt{2}} \left( \, i e^{-i\phi_{SW}}  \ri^\qmg  + \ {\rm h.c.} \
\right) \nonumber \\ 
 \ri&=& \ri^\qmg - {\betab \over
 \sqrt{2}}i\left( \, e^{-i\phi_{SW}}  \rs^\qmg \ + \ e^{i\phi_{SW}} \rl^\qmg
 \,\right) -{i\alphab \over 2} (\ri^\qmg)^{\dagger} \, .
\label{eq:riibwritten}
\end{eqnarray}

The corresponding eigenvalues are corrected by the shifts
$\Gamma_{L,S} \ra \Gamma_{L,S} + \gammab \cdot \dm$, $\gb \ra \gb + \alphab
\cdot\dm$ and $\dm\ra\dm\cdot (1-(\betab /8)^2)$.  
  
The major effect of violation of \qm\ is embodied in the eigenmodes
$\rl$, $\rs$, $\ri$.  These density matrices are no longer pure
density matrices in contrast to their quantum mechanical counterparts
(labelled with a diamond). This loss of purity alters the decay
properties of the beam. For example, the properties of the beam at
large time, ${\tau \gg 1/\gs}$, are dominated by the properties of
$\rl$. The second term on the RHS of the equation for $\rl$ as given
in Eq.~(\ref{eq:riibwritten}) is proportional to $\rs^\qmg$ and is even 
under \cp\ conjugation.  That results in an enhancement of the rate of 
decay into two pions at late time in the evolution of the beam, 
proportional to ${\gammab \over 2}$.  A similar argument leads to 
expect an enhancement by an amount $\propto \betab \,\cos\phi_{SW}$ 
and $\betab \,\sin\phi_{SW}$ in the intermediate time region, $ \tau 
\sim 1/\gb$.  Furthermore, these effects distinguish between $\ko$ and 
$\kob$ (as seen in an appropriate basis) and, consequently, violate 
\cpt\ symmetry.


A proper method of extraction of the parameters $|\Delta|$, $\alphab$,
$\betab$ and $\gammab$ is a method which accounts for possible
correlations or accidental cancelations among these parameters.  Such
a method was first proposed in Ref.~\cite{hp}, the essence of which is
as follow.  $(1)$ $\beta$ and $\gamma$ are extracted by comparing
measurements of the $2 \pi$ decay rates $R_L$ and
$\etapp=|\etapp|\exp(i\phipp)$ along with the semileptonic asymmetry
$\delta_L$.  In \qm, these quantities relate according to $ R_L =
|\etapp|^2$ and ${\delta_L / 2} = \Re\, \etapp$.  After allowance has
been made for violation of \qm, they relate according to $ R_L \simeq
|\etapp|^2 + \gammab/2 + 2 \sqrt{2}\betab \,|\etapp| $ and $
{\delta_L/ 2} = \Re\, \etapp - \sqrt{2}\,\betab \, \sin \phi_{SW}$.
The geometry of these corrections is given in Ref.~\cite{hp}.  $(2)$ A
linear combination of $\beta$ and $|\Delta|$ is constrained from the
value of $\phipp - \phi_{\rm SW} = 0.21 \pm 0.6$ (PDB'96) by
accounting for the dominance of the isospin-0 $\pi\pi$ decay
channel. $(3)$ Finally, $\alphab$ is obtained from the shift $\gb \ra
\gb + \alphab \cdot \dm$ mentioned above.  This yields:\footnote{ -
These bounds depart slightly from the ones obtained in Ref.~\cite{hp}.
This reflects a recent adjustment of the value of $\phipp$ given in
the PDB'96.\newline - In obtaining these bounds, we set to zero the
\cpt\ violating \qm\ perturbations of the decay amplitudes in the
two-pion and semi-leptonic channels.  For more complete formulas, see
\cite{hp}.} \, ${\alphab} \, \le \, 10^{-2}$, $\, \betab \, = \, (
-0.2 \pm 0.67 ) \times 10^{-4}$, $\, \gammab \, = \, ( 0.3 \pm 0.54)
\times 10^{-6}$ and $\, \betab/\sqrt{2} \pm |\Delta| \, = \, ( 0.08
\pm 0.24) \times 10^{-4}$.  These bounds are in agreement with the
ones obtained in Ref.~\cite{ellisetal} under the
assumption that $\Delta=0$.

To answer a question asked during this workshop, neither of 
these parameters contributes more than $10\%$ of the total \cp\ 
violation observed in the \kkbar\ system.

We conclude this section by noting that, these bounds are of the 
order of the expectations presented earlier on the assumption that 
\cptv\ arises from Planck scale physics.  This is an incentive to 
pursue the quest of \cptv\ in the kaon system.

 \section*{Tests of \qm\ at a $\phi$-factory}

 At a \phif, a spin-1 meson decays to an antisymmetric state of two
kaons which propagate with opposite momenta.  If the kaons are
neutral, the resulting wavefunction, in the basis of \cp\ eigenstates
$\ket{K_1}$, $\ket{K_2}$, is $\phi\rightarrow \bigl(
|K_1,p>\otimes|K_2,-p>- |K_2,p>\otimes|K_1,-p> \bigr){\sqrt 2}$. The
two-kaon density matrix resulting from this decay is a $4 \times 4$
matrix $P$, which, in the context of generalized \qm, evolves
according to Eq.~(\ref{eq:densitytime}). When expressed in terms of
the eigenmodes $\rl$, $\rs$ and $\ri$, it takes the form${}^{6]}$
\begin{eqnarray} P &=& \half \biggl[ \rs \otimes \rl + \rl \otimes \rs
- \ri\otimes \ri^{\dagger} -\ri^{\dagger} \otimes \ri \biggr]\nonumber
\\ &+& \biggl[ \ {\betab\over \sqrt{2}}i \left(\, ( e^{i\phi_{SW}}\rs +
e^{-i\phi_{SW}}\rl)\otimes \ri + \ri \otimes ( e^{i\phi_{SW}}\rs +
e^{-i\phi_{SW}}\rl)\, \right) \nonumber \\ &+& \alphab \, i \,
\ri\otimes \ri + {\gammab\over 2} (\, \rl\otimes \rl\, -\, \rs\otimes
\rs \, ) \ + \ {\rm h.c.} \  \biggl].
\label{eq:Pform}
 \end{eqnarray} The time dependence of each term is obtained from the
substitutions $\rho_i \otimes \rho_j\ra$\break $\rho_i \otimes \rho_j
exp(-\lambda_i\tau_1-\lambda_j\tau_2)$ with $\lambda_L=\gl$,
$\lambda_S=\gs$ and $\lambda_i= \gb+i\dm$.  The first line (inside the
first pair of brackets) has the canonical form predicted by \qm\ after
the replacement $\rk^{\qmg} \ra \rk$, while the remaining terms
(inside the second pair of brackets) have a peculiar dependence on
$\tau_1$ and $\tau_2$ and provide an unambiguous method to isolate the
$\dh^{ij}$ parameters from the \qm\ \cpt\ violating
perturbations.\footnote{That is, from the parameter $\Delta$ a well as
other \cpt\ violating parameters in the decay amplitudes.} The basic
observables computed from $P$ are double differential decay rates,
$\P(f_1,\tau_1; f_2,\tau_2)$, the probabilities that the kaon with
momentum $p$ decays into the final state $f_1$ at proper time $\tau_1$
while the kaon with momentum $(-p)$ decays to the final state $f_2$ at
proper time $\tau_2$.  As an illustration, let us consider the decay
into two identical final states $f_1=f_2=f$.  In \qm, the quantity
$\P(f,\tau_1; f,\tau_2)$ has only an overall dependence on the choice
of the final state $f$: its dependence on the two times $\tau_1$,
$\tau_2$ is entirely
fixed by the initial antisymmetry of the wave function which is
preserved by the Hamiltonian evolution.  This characteristic is lost
when violation of quantum mechanics is incorporated as in
Eq.~(\ref{eq:Pform}).  One can, for instance, interpolate the double
decay rates into identical final states $\P(f,\tau_1;f,\tau_2)$ on the
line of equal time $\tau_1=\tau_2$.  This quantity vanishes
identically according to the principles of \qm\ and is thus of order
$\alphab$, $\betab$ and $\gammab$.  As an illustration, the
semileptonic double decay rate at equal time yields($\ell^\pm \equiv
\pi^\mp\ell^\pm\nu$)
\begin{eqnarray}
\P(\ell^\pm,\tau;\ell^\pm,\tau) &/& \P(\ell^\pm,\tau;\ell^\mp,\tau)\,  =
\, {1\over2}\bigl[1- e^{-2(\alphab-\gammab)\dm\tau}\bigl(1-\alphab \,
\sin 2\dm\tau\bigr)\bigr] + {\gammab\over 2}\sinh(\dg\tau)\nonumber \\ &\pm&
\,2\,\sqrt{2}\, \betab \, \bigl[\sin(\dm\tau-\phi_{SW})e^{-\dg\tau/2}
+ \sin( \dm\tau+\phi_{SW})e^{+\dg\tau/2}\bigr]\, .
\label{eq:Qratll}
\end{eqnarray}   
The three coefficients $\alphab$, $\betab$, and $\gammab$ are selected
by terms with different time dependences.

\section*{Outlook}

One might be concerned that attempts to depart from unitarity may be
plagued with inconsistencies and consequently be ruled out.  This is
indeed a possibility.  For example, the authors of Ref.~\cite{hp}
pointed out that the spurious time-dependence which arises in the
evolution of correlated kaons resulting from the decay of a $\phi$
(Eq.~(\ref{eq:Pform})), and which may signal violation of \qm\ at a
$\phi$-factory, is a consequence of the loss of the initial
antisymmetry of the wave function, otherwise preserved by a unitary
evolution.  This antisymmetry was initially guaranteed by conservation
laws.  Do these non-unitary effects intrinsically violate conservation
laws?\footnote{ It was pointed out in Ref.~\cite{ellishns} that, in
the extended framework given by Eq.~(\ref{eq:densitytime}), violation
of a conservation law no longer implies violation of a symmetry
principle.}  This possibility has been pointed out in Ref.~\cite{bps}
and used to argue the inconsistency of Hawking's idea.  These
arguments are however not full proof as shown, for example, in
Ref.~\cite{sred}.  Two attitudes are then possible: one may await the
resolution of these issues on a theoretical ground or one may go on
and resolve them in a laboratory.  The latter seems to be the natural
path to follow in the instance where existing theories are not able to
address a question whose answer is at the reach of an
experiment.  As an example, let us contemplate for a moment, violation
of angular momentum resulting from loss of unitarity in the simplest
possible system: a spin ${1 \over 2}$ atom in a magnetic field.
Evolving that system with Eq.~(\ref{eq:densitytime}) results in the
relaxation, $\sim exp(-\dh \tau)$, into an equal mixture of spin up and
spin down, explicitly violating angular momentum conservation.
Authors of Ref.~\cite{norval}, in their experiments on atomic electric
dipole moments, have observed a spin ${1 \over 2}$ Xe~129 atom precess
and relax over a period of a few hundred seconds, providing the
constraints $|\dh| \ll 10^{-26}$ GeV. Furthermore,\footnote{I thank
Norval Fortson for bringing these experiments to my attention.}
Helium~3 atoms, used as a gaseous polarized target for use with a
polarized electron beam at SLAC,${}^{11]}$ were observed to precess
during a time larger than $75$ hours, bringing the bound down to $\sim
10^{-31}$ GeV much below our naive expectations, $ > 10^{-19}$ GeV.
One can safely conclude that whether or not it is consistent to
include a term $\dh$ in the evolution of a spin $\half$-system, this
term, if present, ought to be too small to give observable
consequences.  This is a particular instance where experimentalists do
not have to wait for theorists to develop a theoretical framework in
order to probe a fundamental property of Nature.  Similarly, \cptv, if
it originates from Planck scale physics, may be at the reach of future
kaon experiments and a genuine phenomenological parameterization in
terms of $\{\Delta, \dh^{ij}/\dm \}$ may be all what is needed to
discover it or demonstrate its irrelevance to low energy physics.

Work supported by the Department of Energy, contract DE-FG03-96ER40956.

\end{document}